\documentclass[%
 aip,
 amsmath,amssymb,
reprint,%
]{revtex4-1}
\draft % marks overfull lines with a black rule on the right

\usepackage{graphicx}% Include figure files
\usepackage{dcolumn}% Align table columns on decimal point
\usepackage{bm}% bold math
\usepackage{bbold}

\usepackage[utf8]{inputenc}
\usepackage[T1]{fontenc}
\usepackage{mathptmx}

\usepackage{units}
\usepackage{verbatim}
\usepackage{xcolor}

\def\ifcomment{\iffalse}

\def\pdf{f}
\def\mass{m}
\def\charge{q}

\def\Hamiltonian{H}

\def\Bfield{B}
\def\VBfield{\VB}
\def\bhat{ \mathbf{\hat b}}
\def\Efield{E}
\def\VEfield{\VE}

\def\Eperp{\Efield_\perp}
\def\Epar{\Efield_\|}

\def\kpar{k_\|}
\def\kperp{k_\perp}

\def\plasma{p}
\def\cyc{B}

\def\omgp{\omega_\plasma}

\def\OmgB{\Omega_\cyc}
\def\OmgB{\Omega_\cyc}
\def\elec{E}
\def\OmgE{\Omega_\elec}

 \def\OmgD{\Omega_D}

\def\gyro{\rho}
\def\Vgyro{\boldsymbol  \rho}
\def\gyrophase{\theta}

\def\BesselJ{J}

\def\ncoord{n}
\def\pcoord{p}
\def\xcoord{x}
\def\ycoord{y}
\def\zcoord{z}

\def\vcoord{v}

\def\Rcoord{R}
\def\Vcoord{V}
\def\Xcoord{X}

\def\Vkcoord{\mathbf k}

\def\Vwcoord{\mathbf w}

\def\VRcoord{\mathbf R}
\def\VVcoord{\mathbf V}

\def\vpar{\vcoord_\|}
\def\vperp{\vcoord_\perp}

\def\kcoord{k}
\def\tcoord{t}

\def\gyrophase{\theta}

\def\Phipert{  \Phi_1}

\def\vperp{\vcoord_\perp}

\def\Bz{B_0}

\def\rx{\xcoord}
\def\ry{\ycoord}
\def\rz{ \zcoord}

\def\vx{\vcoord_\rx}

\def\vz{\vcoord_\rz}

\def\px{\pcoord_\rx}
\def\py{\pcoord_\ry}
\def\pz{\pcoord_\rz}

\def\OmgSlab{\Omega_0}
\def\Xcenter{\Xcoord}

\def\Vnabla{\boldsymbol  \nabla}

\def\VA{\mathbf A}
\def\VB{\mathbf B}
\def\VE{\mathbf E}
\def\VF{\mathbf F}
\def\VJ{\mathbf J}
\def\VP{\mathbf P}
\def\VR{\mathbf R}

\def\VV{\mathbf V}

\def\Vp{\mathbf p}
\def\Vu{\mathbf u}
\def\Vr{\mathbf r}
\def\Vv{\mathbf v}
\def\Vw{\mathbf w}

\def\Vpi{\boldsymbol \pi}
\def\Vrho{\boldsymbol  \rho}

\newcommand\norm[1]{\left| #1\right|}
\newcommand\avg[1]{\left< #1 \right>}
\newcommand\tr[1]{\mathrm{tr}\,{#1}}

\def\CA{\mathcal A}

\def\CO{\mathcal O}

\def\Jmu{J}
\def\xhat{\mathbf{\hat e}_x}
\def\yhat{\mathbf{\hat e}_y}

\def\ehat{\mathbf{\hat e}}

\def\gyromag{a}
\def\Vgyromag{\mathbf a}
\def\wmag{u}
\def\Vwmag{\Vu}
 
\def\krho{c} 
\def\kdotrho{\krho_k}  
  
\def\density{n}

\def\vpar{v_\|}

 \def\Density{N}

\def\pressgc{{P}}
\def\pressgctensor{ \overleftrightarrow{P}}

\def\magnetization{\mu}
\def\Vmagnetization{{\boldsymbol \magnetization}}
\def\Magnetization{M}
\def\VMagnetization{{\mathbf  \Magnetization}}
\def\polarization{\pi}
\def\Vpolarization{{\boldsymbol\polarization}}
\def\Polarization{P}
\def\VPolarization{{\mathbf \Polarization}}
\def\Pdf{F}

\def\magflux{\Psi}

\def\vdrift{V}
\def\Vvdrift{\VV}
\def\vdrift{V}
\def\Vvdrift{\VV}
\def\Action{\CA}

\def\Hessian{\hat \Psi}

\def\Vp{{\bf p}}
\def\VZ{{\bf Z}}
\def\Mmatrix{{\hat M}}
\def\zeromatrix{{\mathbb 0}} %{{\mathds 0}} %{{\mathbbm 0}} 
\def\unitmatrix{{\mathbb 1}} %{{\mathds 1}} %{{\mathbbm 1}}
\def\symplecticmatrix{{\mathbb J}} %{{\mathds J}} %{{\mathbbm J}}

\def\Bz{B_{0,z}}

\def\DelPotential{\Psi''_0}

\def\ddt{\tfrac{d}{dt}}

\def\VPDrift{\VP_{0, \mathrm{gc}} }
\def\HamDrift{H_{0, \mathrm{gc}} }
\def\HamOsc{H_{0, \mathrm{osc}}}
\def\HamGK{H_{1, \mathrm{gk}}}

\begin{document}

\preprint{LLNL-JRNL-813195}

\title[Guiding Center \& Gyrokinetic Orbit Theory for 
Large Electric Field Gradients]{Guiding Center and Gyrokinetic Orbit Theory for Large Electric Field Gradients and Strong Shear Flows}
% Force line breaks with \\

\author{Ilon Joseph}
\email[]{joseph5@llnl.gov}
\affiliation{Fusion Energy Sciences Program, Lawrence Livermore National Laboratory, L-440, P.O. Box 808, Livermore, CA  94550-0808, USA}

\date{\today}% It is always \today, today,

\begin{abstract}
The guiding center and gyrokinetic theory of magnetized particle motion is extended to the regime of large electric field gradients perpendicular to the magnetic field.  
A gradient in the electric field directly modifies the oscillation frequency and causes the Larmor orbits to deform from circular to elliptical trajectories.  
In order to retain a good adiabatic invariant, there can only be strong dependence on a single coordinate at lowest order, so that resonances do not generate chaotic motion that destroys the invariant.  
When the gradient across magnetic flux surfaces is dominant, the guiding center drift velocity becomes anisotropic in response to external forces and additional curvature drifts must be included. 
The electric polarization density remains gyrotropic, but both the polarization and magnetization are modified by the change in gyrofrequency.
The  theory can be applied to
shear flows that are even stronger than those observed in the edge transport barrier of a high-performance tokamak (H-mode) pedestal, even if the toroidal field is as small as or even smaller than the poloidal field.
Yet, the theory retains  a mathematical form that is similar to the standard case and can readily be  implemented within existing simulation tools.
\end{abstract} 

\maketitle

%%%%%%%%%%%%%%% Intro %%%%%%%%%%%%%%%%
\section{Introduction \label{sec:introduction}} 
Understanding the motion of charged particles in electromagnetic  fields is an important subject in many fields of physics.   For magnetized particles, the guiding center\cite{Gardner59pr,Kruskal62jmp, Northrop63book, Morozov66rpp, Boozer80pof, White82pof, Littlejohn81pof,Littlejohn83jpp} (GC) and gyrokinetic\cite{Frieman82pof,Lee83pof,Dubin83pof,Brizard89jpp,Hahm96pop} (GK) approaches, which exploit the adiabatic invariance of the magnetic moment, $\mu$, have been developed to a high degree of sophistication (recently reviewed in Refs. \onlinecite{Cary09rmp,Brizard07rmp}). Yet, for a number of applications, one must understand the effects of both strong magnetic fields, $\VB$, and strong electric fields, $\VEfield$.   
Thus, extending GC/GK  theory to the smallest possible guide field would allow application to the widest variety of physical scenarios encountered in both nature and the laboratory.
  
The goal of this work is to derive a simple and useful extension of GC/GK orbit theory to the largest electric field gradient that still maintains a good adiabatic invariant and to clearly explain the  limits of these approximations with regard to  shear flows.
This work only treats the particle orbits, and, a complete derivation of the GC/GK Maxwell's equations still needs to be performed in order to obtain a self-consistent theory.

In fact, there is a simple extension to GC/GK theory  that retains the essential modifications caused by an electric field gradient and  unites the theory with the low-frequency limit of oscillation center theory.\cite{Dewar73pof, Cary77prl, Johnston78prl, Dewar78jpa}
If one expands the Hamiltonian to quadratic order in momenta and coordinates, then the electric field  gradient directly modifies the oscillation frequency, causes the charged particle orbits  in the perpendicular plane to deform from circular to elliptical trajectories, and generates an anisotropic drift motion.  
 While similar ``orbit-squeezing'' effects have been accounted for in the GC drift motion,\cite{Hazeltine89pfb,Shaing92pfb,Kagan08ppcf} they have not been accounted for in the gyromotion itself. 
Yet, these simple modifications provide the accuracy needed to address the  large gradient regime and can readily be implemented within standard GC/GK simulation tools.

For magnetized plasmas, strong electric field gradients can occur at the interface between regions where the dominant nonambipolar transport processes change and, hence, require a change in the ambipolar electric field. 
This typically occurs across the last closed flux surface of a magnetic confinement device where field lines transition from being closed to openly contacting material surfaces, and clearly occurs within the plasma sheath itself. 
This can also potentially occur across a  separatrix in the magnetic flux function that generates magnetic islands or at the interface between regions of chaotic and integrable magnetic field lines.

When large radial electric field gradients occur in a closed field line region, the strongly sheared flow can suppress  turbulence and form a ``transport barrier'' where  gradients in density and temperature also become large.  
The smallest observed scale lengths  occur for edge transport barriers in the pedestal and scrape-off layer  (SOL) of a  high-performance (H-mode) tokamak, where  shear flows are strong.  
In both the pedestal \cite{Wagner82prl, Burrell94pop, Wagner07ppcf, Kagan08ppcf}  and the SOL, \cite{Eich11prl,Goldston12nf, Scarabosio13jnm, Eich13nf, Sun15ppcf} observations and predictions of the edge scale lengths can be similar in size to the poloidal gyroradius.  
 
Although the poloidal gyroradius  is larger than the toroidal gyroradius by the ratio of toroidal to poloidal field strength, the formal separation of scales needed for standard GC theory no longer exists. \cite{Sturdevant17pop} 
For example, one might estimate the corrections to scale as the ratio of poloidal to toroidal field squared, $(B_p/B_t)^2\sim 0.01$, or even worse, as one over the safety factor squared, $q_{mag}^{-2}=(RB_p/aB_t)^2\sim 0.1$, where $R$ and $a$ are the major and minor radii.
While a subsidiary expansion could be performed, \cite{Hahm09pop} such approximations would not apply to configurations where the poloidal and toroidal fields are comparable, such as spherical tokamaks, spheromaks, and reversed field pinches (RFP's). 
Moreover, this suggests that in configurations without toroidal/guide field, such as field-reversed configurations (FRC's) and Z-pinches, edge scale lengths could potentially reach the order of the gyroradius itself. If this is the case, then standard GC/GK theory would no longer be valid in the edge region. 

Extending GC/GK theory to smaller guide field would also be useful in diverse applications such as accelerator and beam physics, astrophysical and space plasmas, industrial plasmas, sheath physics, laser-plasma interactions, and high energy density science. 
For example,  GC/GK theory has  been applied to plasma turbulence in astrophysical scenarios \cite{Howes06apj, Schekochihin09apj} that rely on a static mean-field approximation for $\VB$. However, astrophysical magnetic fields are often topologically nontrivial and  have regions where the magnetic field must vanish.  Again, it would be desirable to extend GC/GK theory to the smallest possible guide field in order to treat the largest possible domain.   
 
The equations of motion for a nonrelativistic particle of mass $\mass$ and charge $\charge$ are
\begin{align} \label{eq:charge_particle}
\dot \Vr &= \Vv & \mass \dot \Vv &= \charge (\VE+\Vv\times \VB).
\end{align}
(For simplicity, only the non-relativistic case will be discussed.)
It is well known that a perpendicular electric field leads to a perpendicular $\VE\times \VBfield$ drift velocity, but a constant drift velocity can always be removed by a change of reference frame. Thus, it is only the variation of the fields that causes an obstruction to integrability.  
The electric field corrections become important when the electric oscillation frequency 
\begin{align}\label{eq:omge}
\OmgE=(-\charge \Vnabla \cdot \VEfield_\perp/ \mass)^{1/2}
\end{align}
 becomes as strong as the cyclotron frequency 
 \begin{align}
 \OmgB=\charge \Bfield/\mass.
 \end{align}
The electric oscillation frequency is simply the frequency associated with simple harmonic motion at the bottom of an electric potential well.
If the magnetic field has weak spatial variation, then it is related to the parallel component of the curl of the $E\times B$ velocity via $\OmgE^2/\OmgB \simeq \bhat \cdot \nabla\times \VV_E$.  
Note that, for a quasineutral system, $\OmgE$ is always much smaller than the plasma frequency, $\omgp$, since  $(\OmgE/\omgp)^2=-\varrho/\charge \density$ is the ratio of charge density $\varrho$ to particle density $\density$. 

Because the electric oscillation frequency, $\OmgE$, has a weaker dependence on the charge to mass ratio, $\charge/\mass$, than the cyclotron frequency, $\OmgB$, the effects are largest for weakly charged massive particles.
For a 1 \unit{T} magnetic field, the large shear condition, $ \Efield'/\Bfield \sim \OmgB$, occurs at 176 $\unit{GV/m^2}$ for electrons, 95.8 $\unit{MV/m^2}$ for protons,  and $\sim  1$~$\unit{MV/m^2}$ for a singly-charged heavy ($A\sim 100$) impurity ion.   
 The effects are clearly important for weak magnetic fields and sufficiently low temperatures for partially ionized impurity ions and molecules to exist.
 It could also apply to sufficiently small dust grains that have a large enough charge to mass ratio to be considered magnetized.
 Such conditions can be found in diverse application areas such as industrial, astrophysical, and space plasmas, as well as in confinement configurations such as FRCs and RFPs.

Present day tokamak H-mode pedestals typically have a large electric field shear on the order of $\sim 1-10$ $\unit{MV/m^2}$ and temperatures on the order of $\sim 1-10$~\unit{keV}. \cite{Burrell94pop, Wagner07ppcf, McDermott09pop, Leonard14pop}
The ratio of electric to magnetic oscillation frequency is typically of order $1-2\%$ for the main ion species and could be as large as $\sim5-10$\% for partially ionized heavy impurities such as tungsten. 
If the  electric field shear were to become $10 \times$ larger, then the effects would become significant because this would imply $\sim10-20$\% corrections for main ions and $\sim\CO(1)$ corrections for impurities.
Yet, the theory developed here would still be applicable in such extreme scenarios.

In the next section, the constraints of adiabatic invariance will be discussed.  
These constraints imply that the fields must only depend on a single coordinate, e.g. the magnetic flux function, to lowest order.  
The  guiding center ordering assumptions are relaxed to the largest possible perpendicular electric field gradient in Sec.~\ref{sec:orderings}.
A natural and straightforward approach to the derivation of the guiding center drift and Hamiltonian is outlined in Sec.~\ref{sec:GC_derivation}.
In order to clearly explain the issues involved when more than one oscillation frequency is large, 
 the case of strong 2D variation near a fixed point in the electric potential is considered in Sec.~\ref{sec:2d_variation}.
Then, Sec.~\ref{sec:1d_variation} discusses the useful approximation that only the 1D variation across magnetic flux surfaces is dominant.
In this case, the Larmor orbits become elliptical (Sec.~\ref{sec:1d_GC_osc}) and the drifts (Sec.~\ref{sec:1d_GC_drift}) become anisotropic in response to external forces. 
Section~\ref{sec:1d_GC_pol} describes the modification to the linear GC polarization and magnetization response.  
Extension to gyrokinetic orbit theory for strong 1D variation is briefly described in Sec.~\ref{sec:GK}.  
The conclusions are summarized in the final section.

%%%%%%%%%%%%%%% Adiabatic %%%%%%%%%%%%%%%%
\section{Adiabatic Invariance \label{sec:adiabatic}} 
When the electric field is as strong as the magnetic field, then no adiabatic invariants will exist unless there are approximate symmetries that are valid at lowest order. 
If $\VEfield$ depends on all of the coordinates,  then the motion for any particular trajectory will only be cyclic if there is an effective  potential well in all directions and the particle drift velocity will be determined by the location of the center of the well. 
However, if there is no scale separation, the three frequencies will be similar in magnitude. In this case, resonances will generate chaotic motion at the first order of perturbation theory and will generically induce Arnold diffusion. \cite{Chirikov79pr, Lichtenberg92book}
 
 For $d+1$ space-time dimensions, there must be $d$  symmetries in order for the trajectories to be integrable. \cite{Lichtenberg92book}
 When the Hamiltonian only depends strongly on a single coordinate $\psi$, there are $d-1$ conserved momenta and time-independence implies conservation of the Hamiltonian. 
 These constraints only allow a strong dependence of the fields and metric tensor on  a single coordinate,  $\psi(\Vr,\tcoord)$, at lowest order. 
 This allows one to solve for the velocity $v^\psi$ and canonical momentum $p_\psi$. Hence one can determine the oscillation period  $2\pi/\Omega = \oint d\psi/2\pi v^\psi $ and the adiabatic invariant  $\Jmu=\oint  p_\psi d\psi/2\pi$.  
 The frequency and shape of the trajectory must be determined numerically for each initial condition and averages over the oscillation period must be computed numerically for each orbit. 
 While computationally expensive relative to GC theory, it is necessary for arbitrarily strong field variation. 
 
For example, strong electric fields  develop  within the  plasma sheath that forms near material surfaces. The geometry commonly employed to mitigate heat fluxes uses material surfaces that have a shallow angle of incidence with respect to magnetic field lines. In this case, the sheath  is nearly perpendicular to the magnetic field lines and has a much weaker parallel component. \cite{Chodura82pof,Cohen98pop,Geraldini18ppcf} 
As first shown by R.~H.~Cohen and D.~D.~Ryutov in Ref. \onlinecite{Cohen98pop}, the adiabatic assumption applies here and can be used to define GC orbit theory in the sheath. 
However, if the Debye length is smaller than the gyroradius, then this requires the approach described in the preceding paragraph rather than the quadratic approximation described in the following. 

 The symmetry constraints only allow a strong dependence of the fields and metric tensor on  a single coordinate,  $\psi(\Vr,\tcoord)$, at lowest order.
 If the electric scalar potential $\Phi$ and vector potential  $\VA$ only depend strongly on space-like coordinate $\psi$ to lowest order, then the fields must satisfy
\begin{align}
\VE & =-  \Vnabla \psi \partial_\psi \Phi - \partial_\tcoord \psi \partial_\psi  \VA 
&
 \VB &=\Vnabla \psi \times  \partial_\psi  \VA   .
 \label{eq:mag_field}
\end{align}
Hence,  to lowest order,  $\psi$ must be a constant of the magnetic field line trajectories, $\VB\cdot \Vnabla \psi =0$, and the field evolution must be ideal \cite{Boozer04rmp} due to the fact that $\VE\cdot \VB=0$.  This implies that there is reference frame in which $\VE=-\Vnabla\Phi$ and that the topology of the field lines is fixed in time. 
Any magnetic field can be locally written in the  Clebsch form $\VB=\Vnabla \magflux \times \Vnabla \alpha$, and, in a topologically toroidal region, $\magflux$ can be taken to be a  single-valued  magnetic flux coordinate indicating the ``radial'' direction. \cite{Boozer04rmp} However, $\alpha$ is generically a multi-valued function, even for integrable field line trajectories. Hence,  only   $\psi(\magflux)$ will be a  globally well-defined coordinate suitable for describing particle motion over a global region; this choice implies $\Vnabla \alpha= \partial_\magflux \VA$.

\section{Guiding Center Ordering Assumptions \label{sec:orderings}}
\subsection{Perturbation Theory}
It is desirable to develop a perturbation theory that is as accurate as possible at lowest order. 
This is because the theory of adiabatic invariants only provides an asymptotic approximation to the exact particle trajectories. \cite{Kruskal62jmp, Lichtenberg92book}
Hence, the asymptotic series will eventually diverge and is not carried to high order in practice.  
In fact, to this day, only a few codes\cite{Belova03pop} have attempted implementing a complete treatment of 2nd order GC effects.   
Methods for achieving superconvergent asymptotic expansions  \cite{Kolmogorov54dak,Chirikov79pr,Lichtenberg92book} are based on determination of the best approximation to the frequency for each iteration. 
This is equivalent to the resummation of an infinite number of terms in the perturbation series. 
Thus, it is especially important to employ an accurate approximation to the frequency of cyclic orbits at lowest order.   

\subsection{The small parameter: $\epsilon\ll1 $}
The derivation of the GC drift equations is based on the assumption that the temporal and spatial scales for variation of the magnetic field $\VBfield$ and electric field $\VEfield$ are longer than the oscillation period $2\pi/\Omega$ and oscillation radius $\gyro$.  Hence, the restrictions for any characteristic frequency $\omega$ and wavenumber $k$ are 
\begin{align}\label{eq:order}
\omega/\Omega \sim \kcoord \gyro \leq \CO(\epsilon) \ll 1
\end{align}
where $\epsilon$ is a small ordering parameter. 
For example, the magnetic field variations must  be assumed to be small $\gyro \nabla \Bfield/\Bfield \sim \partial_\tcoord \Bfield/\Bfield \Omega \lesssim \CO(\epsilon)$ in order to develop a valid GC perturbation theory.
Because the metric tensor and connection appear in the equations of motion, they  must satisfy similar restrictions when expressed in the coordinates used for the calculation.

\subsection{Standard Ordering: $\OmgE/\OmgB\sim\CO(\epsilon^2)$}
The standard GC drift ordering in Refs. \onlinecite{Morozov66rpp, Boozer80pof, White82pof, Littlejohn83jpp, Dubin83pof}  assumes that the electric force is smaller than the magnetic force: $\Efield/\vperp \Bfield \lesssim \CO(\epsilon)$, and, hence, that the electric drift velocity is smaller than the gyration velocity. 
Thus, the electric field shear  is restricted to satisfy $\nabla \Efield/  \OmgB\Bfield \lesssim \CO(\epsilon^2)$.   Unless a change of reference frame is employed, trajectories with $\vperp<\Efield_\perp/\Bfield$ are  excluded.  

\subsection{Large Flow Ordering: $\OmgE/\OmgB\sim\CO(\epsilon)$}
The ``large flow  ordering,''  \cite{Northrop63book, Morozov66rpp, Littlejohn81pof,Wimmel84ps,Hahm96pop,Sugama97pop, Sugama98pop,Hahm09pop} and the recent approach of A. M. Dimits, \cite{Dimits10pop,Dimits12pop} are based on choosing the reference frame with the appropriate drift velocity.  
These theories have  separate conditions for the components perpendicular  $\Efield_\perp/\vperp \Bfield\sim \CO(1)$  and parallel  $\Epar/\vperp \Bfield\sim \CO(\epsilon)$ to the magnetic field.   The stronger restriction on the parallel electric field  is essential for ensuring that the change in the parallel velocity per cycle remains small $  \Delta \vpar /\vperp  \lesssim\CO(\epsilon)$.
For this ordering, the perpendicular electric field shear can be somewhat larger, $\nabla \Efield_\perp/\OmgB \Bfield\lesssim \CO(\epsilon)$, but it must still be small. 

\subsection{Maximal Ordering: $\OmgE/\OmgB\sim\CO(1)$ \label{sec:maximal_ordering}}
Assume that the electric force  variations in the plane perpendicular to  the magnetic field can be large,  $\nabla_\perp  \Efield_\perp/\OmgB \Bfield \sim \CO(1)$, but that  the variations are weaker in the parallel direction: $\Epar/\vperp\Bfield\sim  \kpar /\kperp \lesssim \CO(\epsilon)$.
GC theory still retains a simple form if one assumes that
  \begin{align}
  \nabla_\perp  \Eperp/ \Bfield\OmgB&\sim  \nabla_\perp \nabla_\perp  \Phi/ \Bfield\OmgB \sim \CO(1) 
 \end{align}
but that the higher order gradients are small; i.e. for $n\geq2$
 \begin{align}
 \gyro^n\nabla^{n-1}\Efield /\vperp \Bfield \sim \gyro^n\nabla^n\Phi/\vperp \Bfield\leq \CO(\epsilon).
 \end{align}
One natural way in which this ordering can occur is if all perpendicular gradient scale lengths, $k_\perp$, are small in the sense of satisfying $k_\perp\rho \sim\CO(\epsilon)$. This allows the perpendicular electric field and flow velocity to satisfy $E_\perp/\vperp \Bfield \sim V_\perp/\vperp \sim \CO(\epsilon^{-1})$. 
Because this implies that $\phi/\gyro\vperp \Bfield   \sim \CO(\epsilon^{-2})$, this requires the rather weak parallel scale variation $k_\|\gyro\sim \CO(\epsilon^3)$.

%%%%%%%%%%%%%%% GC Derivation %%%%%%%%%%%%%%%%
\section{Guiding Center Derivation \label{sec:GC_derivation}}
\def\Vgyrophase{\boldsymbol\gyrophase}
\def\VJmu{\mathbf J} 
\subsection{ Lagrangian Approach}
While there are many approaches to deriving the equations of motion, \cite{Littlejohn81pof, Littlejohn83jpp, Brizard95pop} the route that appears to require the least amount of algebra is the approach of Ref.~\onlinecite{Dimits10pop}.  
Consider modifying the charged particle action principle
\begin{align} \label{eq:action_integral}
\Action[\Vr;\tcoord] =\int \left[\tfrac{1}{2} \mass  \dot \Vr^2 +\charge \VA\cdot \dot \Vr -\charge \Phi \right]d\tcoord
\end{align}
 by assuming thatone can express the trajectory as the sum of two parts: one that varies on slow timescales, $\VR$, and one that varies on fast timescales, $\Vrho$, so that
\begin{align}
\Vr&=\VR+\rho^i \ehat_i &
\dot\Vr&=\dot \VR+\dot \Vrho.
\end{align}
%where the basis unit vectors $\ehat_i(\Vr,\tcoord) $  and the guiding center velocity   $\dot \VR=\Vvdrift(\Vr,\tcoord)$ are  functions of $\Vr$ and $\tcoord$. 
The separation into slow and fast timescales is accomplished by defining a special coordinate system that provides this separation. 
Assume that are there are  phase coordinates, $\Vgyrophase=\{\gyrophase^i\}$, that encode the fast time dependence and canonically conjugate adiabatic invariant coordinates, $\VJmu=\{\Jmu_i\}$, that are well-conserved.  
The guiding center position, $\VR$, evolves slowly, so it must be independent of the fast phases and can also be used as slow coordinates in addition to the $\VJmu$.
Hence, the basis unit vectors, $\ehat_i( \tcoord, \VR, \VJmu )$, that define the direction of $\Vgyro$ and the guiding center velocity, $\dot \VR:=\Vvdrift(\tcoord, \VR, \VJmu)$, are assumed to be functions of the slow coordinates, $\VR$, the adiabatic invariants, $\VJmu$, and are only allowed to have a weak $\CO(\epsilon)$ dependence on time, $\tcoord$.
In addition to this same dependence, the components of the gyroradius, $\Vgyro^i(\tcoord, \VR, \VJmu, \Vgyrophase )$,  are assumed to depend on the fast phases, $\Vgyrophase$, in an essential oscillatory manner.

\subsection{Drift Motion}
In the limit of vanishing gyroradius, the GC equations of motion are the usual charged particle equations drifting in the non-inertial reference frame with velocity $\Vvdrift$:
\begin{align}\label{eq:Vdrift} 
 \dot \Vvdrift= \partial_\tcoord \Vvdrift +\Vvdrift\cdot \Vnabla \Vvdrift  &=  \charge \left( \VE+\Vvdrift \times \VB\right) / \mass.
\end{align}
In the equations that follow, the time derivative in the GC reference frame is denoted by  $\dot s:=ds/dt=(\partial_\tcoord+\VV\cdot \nabla)s$ for any quantity $s$. 
The zeroth order GC equation of motion can also be written as
 \begin{align} 
\partial_\tcoord \VPDrift+ \Vnabla \HamDrift &=    \Vvdrift \times  \left(\Vnabla\times \VPDrift \right)
\end{align}
where,  to zeroth order in gyroradius, the Hamiltonian and canonical momentum are 
\begin{align}
\HamDrift&=\tfrac{1}{2}\mass \vdrift^2+\charge\Phi & \VPDrift=\mass\Vvdrift+\charge \VA.
\end{align} 
The assumption that $\partial_\tcoord/\Omega\sim\CO(\epsilon)$ implies that $\HamDrift$ is conserved along a streamline, $\VV\cdot \Vnabla \HamDrift =0$,   to $\CO(\epsilon)$.
Note that, in certain works, such as Ref.~\onlinecite{Cary09rmp}, the term $\mass \Vvdrift_\perp$ is not retained in the momentum, and, in that case, the zeroth order Hamiltonian would have the opposite sign for the kinetic term. 

These equations are just as difficult to solve as Eq.~\ref{eq:charge_particle}, if not more so, because this equation is to be interpreted as a PDE and the initial conditions for $\Vvdrift$ need to be chosen to eliminate  fast oscillatory motion. 
For cases with symmetry, it is potentially possible to determine useful ans\"atze for solving this equations. 
However this is challenging for the general case, because the solution will typically have fast oscillatory motion unless it is pinned to a fixed point where $\VE=0$. 

One natural simplifying assumption is that the convective derivative is also small, so that  the entire LHS, $d\Vvdrift/d\tcoord\sim\CO(\epsilon)$, can be neglected to lowest order. 
In this case, the solution can be determined via an asymptotic expansion that begins with the usual lowest order approximation 
\begin{align}
   \Vvdrift_0 =\vpar \bhat +\VE\times \bhat/\Bfield.
\end{align}

\subsection{Oscillatory Motion}
The trajectories in a neighborhood of the guiding center orbit, $\VR$, are parameterized by the oscillatory term, $\Vgyro$.
If $\Vgyro$ is considered small with respect to the spatial scales in the problem, then the action integral (Eq. \ref{eq:action_integral}) can be expanded in terms of its variational derivatives with respect to $\Vgyro(\tcoord)$. Due to the maximal ordering assumptions of Sec. \ref{sec:maximal_ordering}, contributions up to second order in $\Vgyro$ must be retained in the zeroth order action, denoted $\Action_{0,\mathrm{osc}}$, but higher order terms can be neglected.
Due to the fact that the guiding center orbit satisfies the Euler-Lagrange equations of motion resulting from Eq. \ref{eq:action_integral}, 
the first variation of the action with respect to $\Vgyro(t)$ is a total derivative $d(\Vgyro\cdot\VP_0)$.  
Hence, it is the second variation of the action that determines the oscillatory part of the motion.  

The second variation is derived to be
\begin{multline}
\Action_{0,\mathrm{osc}}[\Vgyro;\tcoord] =\int \left\{\tfrac{1}{2} \mass  \dot \Vgyro ^2 +\Vgyro\cdot \Vnabla\charge \VA\cdot \dot \Vgyro \right.
\\
\left.+ \tfrac{1}{2} \Vgyro\Vgyro:\left[(\Vnabla\Vnabla \charge\VA) \cdot \Vvdrift - \Vnabla\Vnabla \charge\Phi\right] \right\}d\tcoord.
\end{multline}
The Euler-Lagrange equations of motion are
\begin{align} \label{eq:ddotgyro_canonical}
\Vp&=\mass \dot \Vgyro + \Vgyro\cdot \Vnabla \charge \VA
\\
\dot\Vp &= \Vnabla  \charge \VA\cdot \dot  \Vgyro +   \Vgyro\cdot \left[(\Vnabla \Vnabla \charge \VA)\cdot \Vvdrift -\Vnabla \Vnabla \charge \Phi\right].
\label{eq:ddotpgyro_canonical}
\end{align} 
The Euler-Lagrange equations are equivalent to 
\begin{align}  \label{eq:ddotgyro_direct}
 \mass \ddot \Vgyro 
 &=\dot \Vgyro \times\charge\VB+\Vgyro\cdot \left[(\Vnabla \Vnabla \charge \VA)\cdot \Vvdrift -\Vnabla \Vnabla \charge \Phi\right]-\Vgyro\cdot \ddt \Vnabla \charge \VA.  \\
  &=\dot \Vgyro \times\charge \VB +\Vgyro\cdot\left[\Vnabla \charge\VE  -\Vnabla \charge\VB \times \Vvdrift \right].
   \label{eq:ddotgyro_field}
\end{align} 
Inserting the GC drift velocity from Eq.~\ref{eq:Vdrift} yields the equivalent expression
\begin{align}  \label{eq:ddotgyro_flow}
\mass  \ddot \Vgyro =\dot \Vgyro \times \charge\VB+\Vgyro\cdot\left[\Vnabla \mass \dot \Vvdrift -\Vnabla \Vvdrift \times \charge\VB\right] .
\end{align} 
The assumptions imply that the $\partial_\tcoord\Vvdrift$ and $\partial_\tcoord\VA$ terms can be neglected at lowest order. However, the convective derivatives can still make a contribution to $d\Vvdrift/d\tcoord$ and $d\VA/d\tcoord$.

Equations \ref{eq:ddotgyro_canonical}-\ref{eq:ddotpgyro_canonical}, and hence, Eq.~\ref{eq:ddotgyro_direct}, 
are manifestly symplectic. However, for the completely equivalent Eqs.~\ref{eq:ddotgyro_field}-\ref{eq:ddotgyro_flow}, 
the symplectic symmetry is not manifestly apparent because the term $\Vgyro\cdot \Vnabla \dot \VA$ has been split and absorbed into the $\Vgyro\cdot \VE$ and $\Vgyro\cdot \Vnabla \VB\times \Vvdrift$ terms.  
 
The canonical Hamiltonian to second order in gyroradius can be expressed in three 
completely equivalent ways as
\begin{align}
 \HamOsc  &=\tfrac{1}{2} \mass\dot \Vgyro^2 +\tfrac{1}{2} \charge\Vgyro\Vgyro:\left[\Vnabla \Vnabla \Phi- \Vnabla \Vnabla \VA\cdot \Vvdrift\right]\\
 &=\tfrac{1}{2} \mass\dot \Vgyro^2 +\tfrac{1}{2} \charge\Vgyro\Vgyro:\left[\Vnabla \VB\times\Vvdrift- \Vnabla \VE -  \ddt \Vnabla   \VA\right] \\
 &=\tfrac{1}{2} \mass\dot \Vgyro^2 +\tfrac{1}{2}\charge \Vgyro\Vgyro:\left[\Vnabla\Vvdrift\times  \VB- \tfrac{\mass}{\charge} \Vnabla \ddt \Vvdrift -\ddt\Vnabla  \VA \right].
\end{align} 
This last form is equivalent to the perturbative analysis of  Ref.~\onlinecite{Brizard95pop} after neglecting the terms with time derivatives. 
Yet another useful expression for the Hamiltonian can be obtained by taking the inner product of  Eq.~\ref{eq:ddotgyro_field} with $ \Vgyro$ which yields
\begin{align}
  \HamOsc&=    \mass\dot \Vgyro^2 + \tfrac{1}{2}\left[\charge \VBfield\cdot \Vgyro\times\dot \Vgyro -\Vgyro\Vgyro:\ddt \Vnabla\charge \VA  - \mass \tfrac{d}{dt}(\Vgyro\cdot \dot \Vgyro)\right].
\end{align} 
The final term is a total derivative that vanishes for simple harmonic motion and vanishes when averaged over the gyroperiod.  

Since the equations of motion are linear and the Hamiltonian is constant to lowest order, the Hamiltonian
is also equal to an average over the gyroperiods $\HamOsc=\avg{\HamOsc}$ and can be expressed as a sum over adiabatic invariants, $\HamOsc=\sum_i\Jmu_i\Omega_i$, for each independent action invariant $\Jmu_i$.  For example, if there is only a single adiabatic invariant, then $\HamOsc= \Jmu \Omega$.

\def\VG{{\mathbf G}}
\def\CG{{\mathcal G}}
Note that the $\mass \dot \Vgyro^2/2$ term contains the gyrogauge ``space-time'' scalar
\begin{align}
\CG = \Omega  \gyro^j \partial_\gyrophase \Vgyro\cdot \tfrac{d}{dt} \ehat_j =   \Omega \gyro^j \partial_\gyrophase \Vgyro\cdot (\partial_\tcoord+\Vvdrift\cdot \Vnabla) \ehat_j .
\end{align} 
The gyrogauge vector, 
$\VG=\Omega \gyro^j \Vnabla \ehat_j \cdot \partial_\gyrophase  \Vgyro$, first introduced by Littlejohn, \cite{Littlejohn81pof, Littlejohn83jpp} is necessary for handling the gauge freedom available in the definition of the gyrophase.
Usually, the gyrogauge vector is included in the canonical momentum, \cite{Brizard07rmp} so that the form, $d\gyrophase-\VG\cdot d\VR$, is gyrogauge invariant. 
The gyrogauge scalar, 
$G_0= \Omega \gyro^j  \partial_\tcoord \ehat_j\cdot \partial_\gyrophase\Vgyro$,
 is usually neglected in the Hamiltonian  because it is higher order in $\epsilon$.
 In fact, the gyrogauge vector term can also be treated as a correction to the Hamiltonian, $\delta \Hamiltonian = \VG\cdot\VV$, where it combines with the scalar term, to form the space-time scalar, $\CG$, above.\cite{Dimits10pop}

%%%%%%%%%%%%%%% 2D Fixed Point %%%%%%%%%%%%%%%%
\section{Strong 2D Variation  Near A Fixed Point \label{sec:2d_variation} }
If the gradients are strong in $n$ directions, where $n>1$, then there are generically $n$ pairs of independent eigenvalues $\pm\Omega_i$ and associated eigenvectors. 
In this case, a non-oscillatory solution for the drift velocity requires $\Omega^{-1} \VV\cdot\Vnabla\sim \CO(\epsilon)$
which only occurs in the vicinity of a fixed point.
%This implies that the center of the orbital motion is locked to the fixed point location, $\VR \simeq -(\Vnabla\VE)^{-1} \VE $. 
In this case, the center of the orbital motion is locked to the fixed point location, 
which can be determined by expanding the electric field as $0=\VE+\Delta \VR \cdot \Vnabla\VE+\dots$, 
and, hence, $\Delta \VR \simeq  \Hessian^{-1} \VE $, where the matrix $\Hessian$ is defined via $\Hessian_{jk}=-\nabla_kE_j= \nabla_j\nabla_k\Phi$. 
Thus, the GC velocity $\VV=d\VR/d\tcoord $ is controlled by the motion of the fixed point and does not have the same freedom that exists in standard GC theory.

The fields must satisfy certain requirements in order to for the orbits to have even a single conserved adiabatic invariant because there is 
no guarantee that all frequencies, $\pm\Omega_i$, are real.
Since the equations of motion are real, the $\Omega_i$'s generically come in complex conjugate pairs. 
In fact, since the equations of motion are symplectic,
both signs, $\pm\Omega_i$, must appear, and, hence, they generically come in quadruplets $\{\pm \Omega_i,\pm\Omega_i^*\}$.
Moreover, since the frequencies are similar in magnitude, resonances can generate chaotic motion at the first order of perturbation theory and the adiabatic approximation will no longer be valid.
In this case, adiabatic invariants will not be well-conserved for all initial conditions.

Let us examine the orbits close to a fixed point of the electric potential. 
Assuming that the potential, $\Phi$, is a function of magnetic flux $\psi$, then this must also correspond  to an  O-point or X-point in the magnetic field.
Hence, strong gradients can exist in both directions perpendicular to $\VB$, but not the direction parallel to $\VB$.    

The equations of motion in velocity phase space, $\VZ=\{\Vr,\Vv\}$, have the  form $d\VZ/dt=\Mmatrix_{\Vr,\Vv}  \VZ$ where the $4\times4$ matrix, $\Mmatrix_{\Vr,\Vv} $, has the $2\times2$ block form
\begin{align}
\Mmatrix_{\Vr,\Vv} =  \left( \begin{array}{cc}
\zeromatrix & \unitmatrix \\
-\charge\Hessian/\mass &  \symplecticmatrix\OmgB
\end{array}\right)
.
\end{align}
In this expression, $\unitmatrix$ and $\zeromatrix$ are the $2\times2$  unit and zero matrices, respectively, $\symplecticmatrix$ is the $2\times2$ antisymmetric tensor
\begin{align}
\symplecticmatrix=  \left( \begin{array}{cc}
0 &1\\
-1 & 0
\end{array}\right),
\end{align}
and $\Hessian$ is the $2\times 2$ symmetric Hessian matrix  of partial derivatives of the potential in the directions perpendicular to the field, 
\begin{align}
\Hessian_{ij} =   \partial_i\partial_j  \Phi_0-(\partial_i\partial_j  \VA_0)\cdot\Vvdrift_0.
\end{align}
In contrast to the standard GC equations, which are gyrotropic, i.e. isotropic in the plane perpendicular to $\VBfield$, in this case, 
 a specific orthogonal transformation of the basis vectors $\xhat$ and $\yhat$ will diagonalize the $\Hessian$ matrix.  
After applying this coordinate transformation, only the $\Hessian_{xx}$ and $\Hessian_{yy}$ elements are nonzero, which significantly simplifies and illuminates the physical meaning of the results.

The linear equations of motion in canonical phase space $\VZ=\{\Vr,\Vp\}$, with the choice of vector potential, $\VA=\VB\times\Vr/2$, has the $2 \times 2$ block form
\begin{align}
\Mmatrix_{\Vr,\Vp} =  \left( \begin{array}{cc}
 \symplecticmatrix \OmgB/2 & \unitmatrix/\mass \\
-\charge\Hessian - \unitmatrix \mass \OmgB^2/4 &  \symplecticmatrix \OmgB/2
\end{array}\right)
.
\end{align}
Since both forms of the equations of motion are related by a linear similarity transformation, they have the same physical content.

\begin{figure}[t]
\center
\includegraphics[height=2.5in]{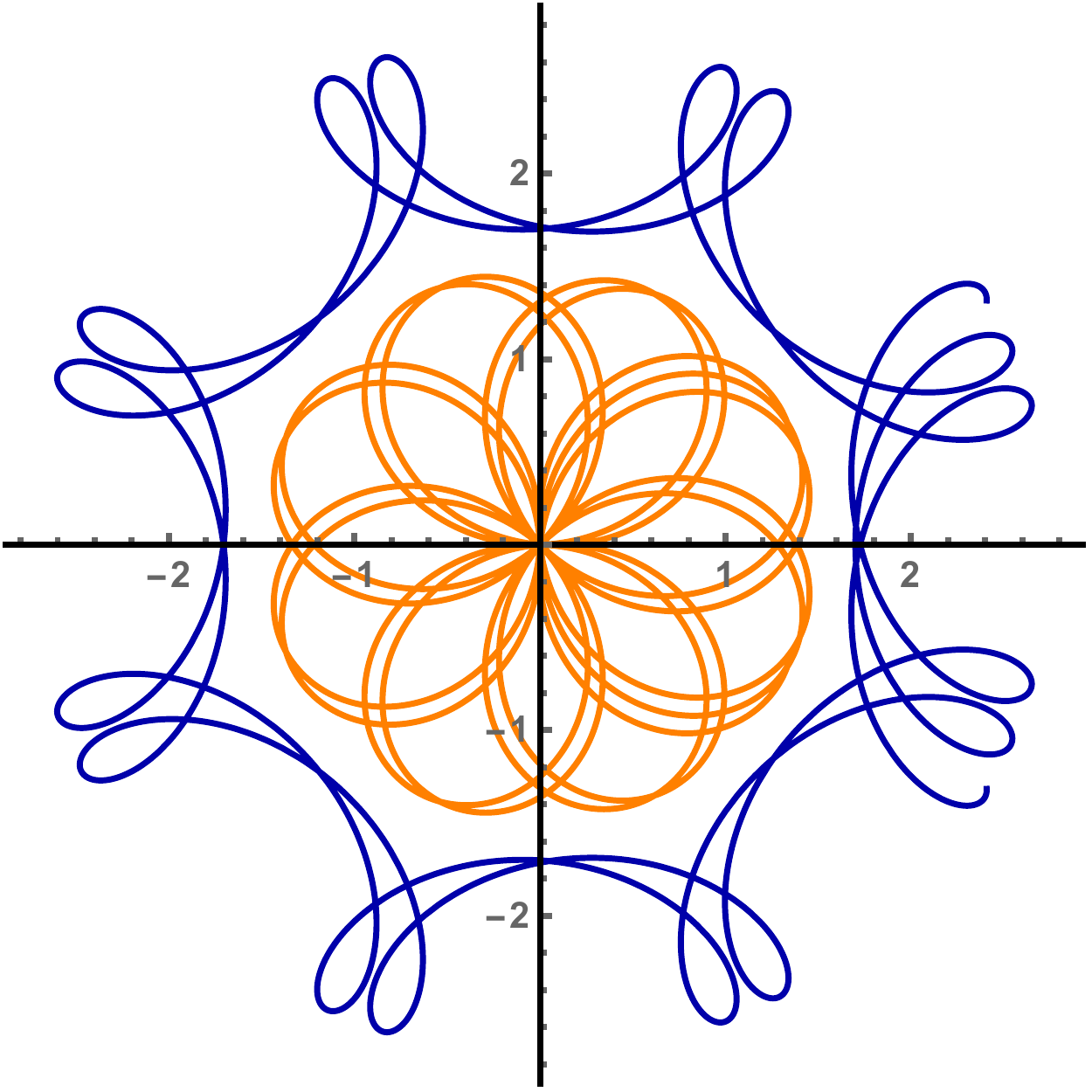}
\caption{Two orbits near a potential  well (O-point)  for the parameters $\Hessian_{xx}/B\OmgB=\Hessian_{yy}/B\OmgB=1/5$, $\OmgE^2/\OmgB^2=2/5$ and $\OmgD^4/\OmgB^4=1/5^2$.}
\label{fig:Orbits-Opoint-Well}
\end{figure}

Because the equations of motion are linear, they have an exact solution. 
 The oscillatory motion in the perpendicular plane  is a linear combination of two ellipses centered at the location of the fixed point, where $\VE=0$. There are two pairs of oscillation frequencies $\pm \Omega_+$ and $\pm \Omega_-$, given by 
 \begin{align} \label{eq:2d_begin}
 \Omega_\pm^2 &=\tfrac{1}{2}\OmgSlab^2 \pm \tfrac{1}{2}\left( \OmgSlab^4-4\OmgD^4\right)^{1/2}\\
 \OmgSlab^2&=\OmgB^2+\OmgE^2
 \label{eq:OmgSlab}
 \\
 \OmgE^2&= \tr{\charge \Hessian/\mass} = \charge \left[\nabla_\perp^2   \Phi_0-(\nabla_\perp^2  \VA_0)\cdot \Vvdrift  \right] /\mass\\
  \OmgD^4&= \det{\charge \Hessian}/\mass
  \label{eq:2d_end}
  .
 \end{align}
After the $\Hessian$ matrix has been diagonalized with the orthogonal transformation described above, the characteristic frequencies simplify to 
 \begin{align} \label{eq:2d_begin}
 \OmgE^2&= \charge(\Hessian_{xx}+\Hessian_{yy})/\mass \\
  \OmgD^4&= \charge^2 \Hessian_{xx}\Hessian_{yy}/\mass^2
  .
 \end{align}
If $\Omega_0^4<4\Omega_D^4$, then both types of motions have an instability.
Hence, in the following, assume that  $\Omega_0^4>4\Omega_D^4$, which implies that at least one of the types of motion is stable.
% Figure~\ref{fig:Orbits-Xpoint} illustrates the case of two trajectories that %are 
 %have stable $\Omega_+^2>0$, but unstable $\Omega^2_-<0$.
%While the theory of adiabatic invariants  applies qualitatively \cite{Best68physica}, as illustrated by  Fig.~\ref{fig:Orbits-Xpoint}, the real difficulty lies in the fact that the quadratic approximation will  break down because the orbit rapidly moves to  a region  that is far from the fixed point where the potentials may have a rather different dependance on space and time.

\begin{figure}[t]
\center
\includegraphics[height=2.5in]{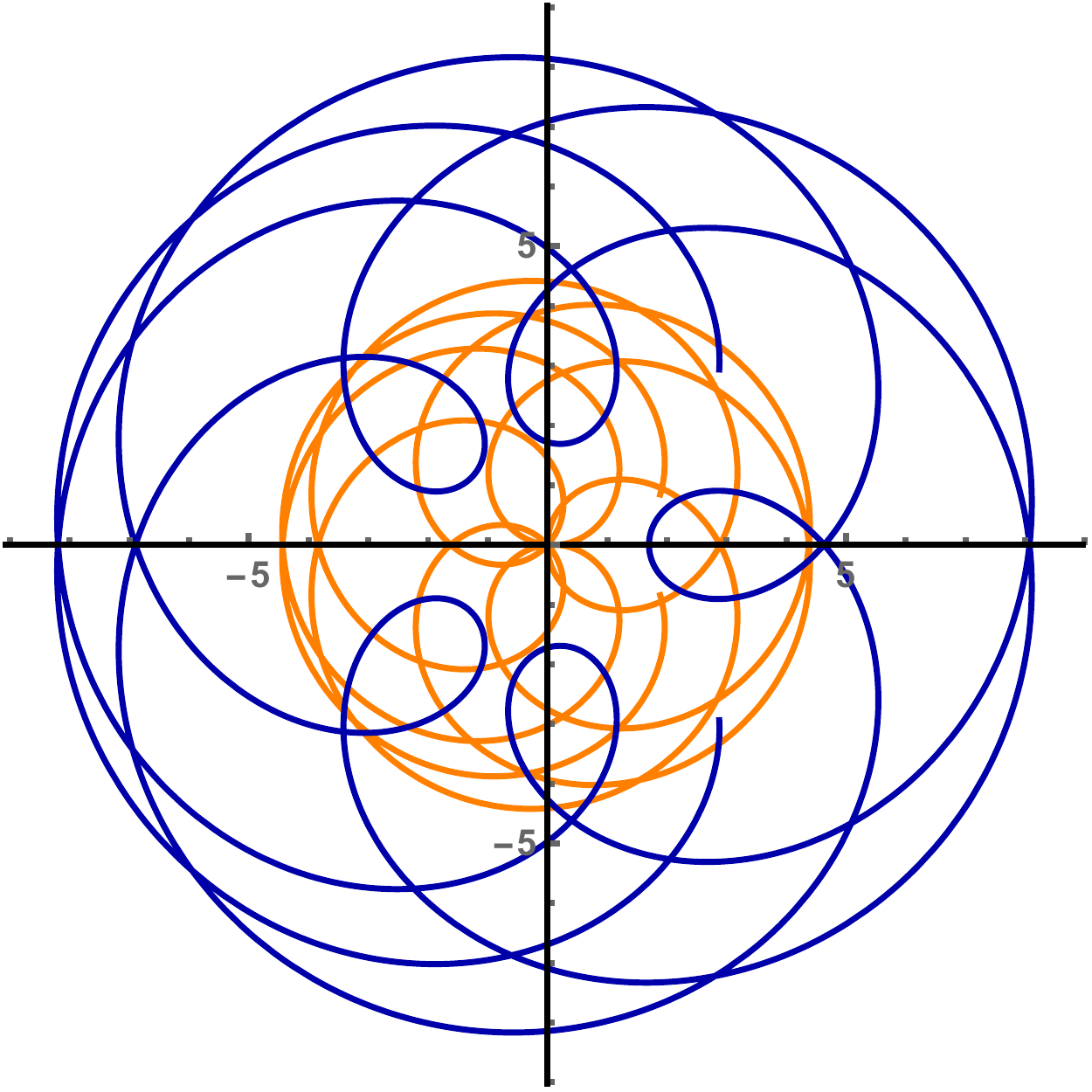}
\caption{ Two orbits near a potential hill (O-point) for the parameters: $\Hessian_{xx}/B\OmgB=\Hessian_{yy}/B\OmgB=-1/5$, $\OmgE^2/\OmgB^2=-2/5$, and $\OmgD^4/\OmgB^4=1/5^2$. }
\label{fig:Orbits-Opoint-Hill}
\end{figure}
If $\Omega_0^2>0$, then both types of motion are stable near a potential O-point, where $\Omega_D^4>0$, and this leads to a theory with two adiabatic invariants, $\Jmu_\pm$.
Figure~\ref{fig:Orbits-Opoint-Well} illustrates the case of trajectories near an electric potential well, while Fig.~\ref{fig:Orbits-Opoint-Hill} illustrates the case of trajectories near an electric potential hill. 
For the case of two stable motions,  $\Omega_+\geq\Omega_-$, and, so, the motion corresponding to $\Omega_+$ can be taken to represent the gyromotion, while the motion corresponding to $\Omega_-$ can be taken to represent a fast drift in the perpendicular plane.
From the point of view of standard GC theory, the %motion corresponding to $\Omega_-^2$ simply represents a very fast drift motion that has been promoted to 
 first adiabatic invariant, $J_+$, corresponds to a modified magnetic moment while the second adiabatic invariant, $J_-$, measures the perpendicular drift around the fixed-point and plays a role similar to the standard parallel ``bounce'' invariant.
Hence, there is no additional freedom for the guiding center to wander from the fixed point.
The lowest order momentum is $\VP=\charge\VA+\mass\VV$ and the lowest order Hamiltonian is 
  \begin{align}
 \Hamiltonian_0 = J_+\Omega_+ + J_-\Omega_- +\mass \Vcoord^2/2 +\charge \avg{\Phi}(J_+,J_-,\ell).
 \end{align}
Here, the potential  $\avg{\Phi}(J_+,J_-,\ell)$ is the orbit-average of $\Phi$  and its only spatial dependence is on the field line length $\ell$.

In the limit where $\OmgB\gg\OmgE$, the magnetic moment is $J_+$ and the gyrofrequency is $\Omega_+\simeq \OmgSlab\simeq \OmgB+\OmgE^2/2\OmgB+\dots$, which agrees with the prediction of the large flow GC ordering. \cite{Brizard95pop, Hahm96pop, Hahm09pop}
In this case, the slower eigenvalue,   $\Omega_{-}^2\simeq -\OmgD^4/\OmgSlab^2$, is comparable to the frequency of the drift motion around the magnetic surface and the entire trajectory must be determined to obtain an accurate approximation to  $J_-$ and $\Omega_-$. 
However, this case can be approximated well by considering 1D variation alone, as discussed in the following section.

\begin{figure}[t]
\center
\includegraphics[height=2.5in]{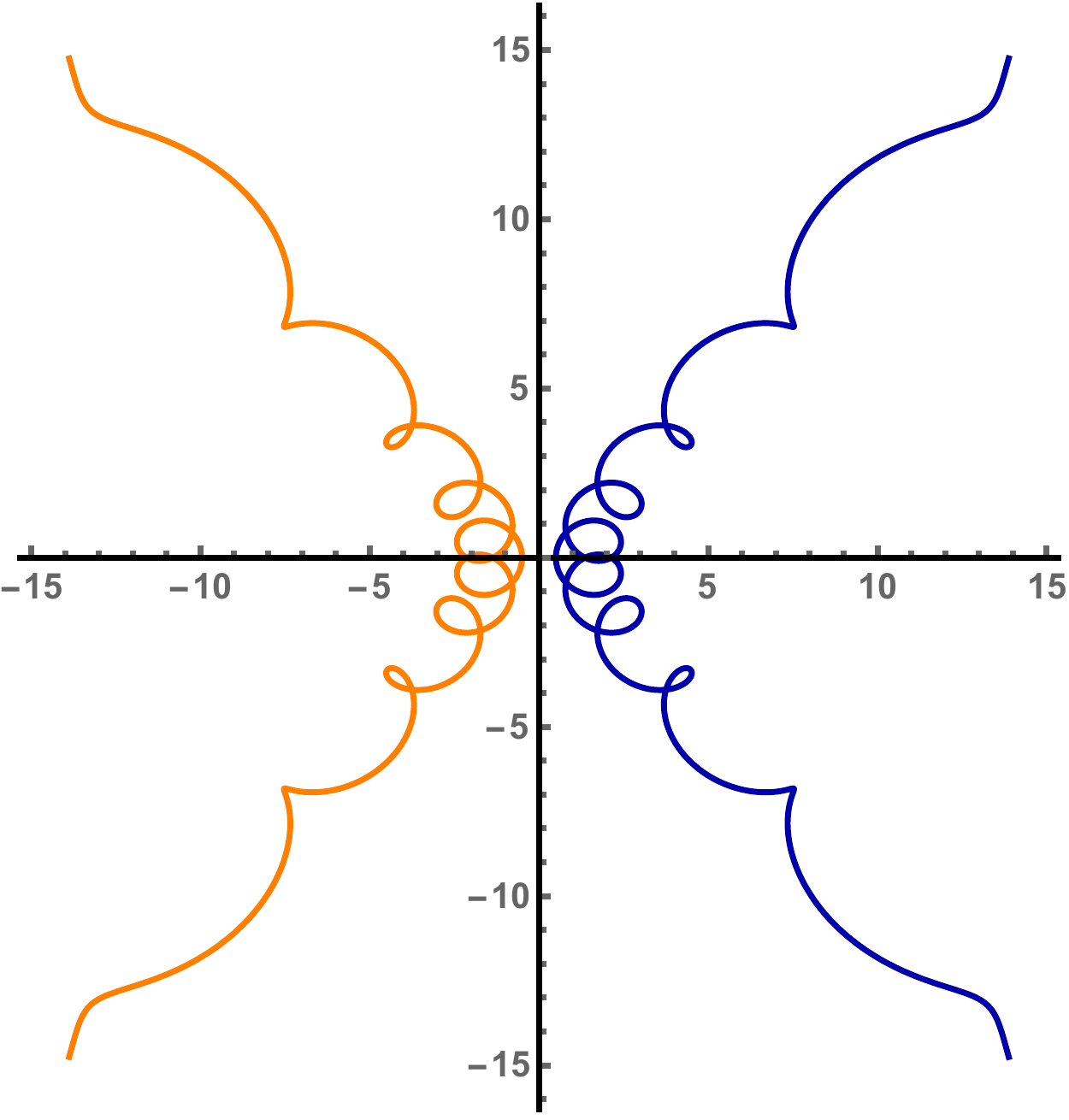}
\caption{ Two orbits near a potential X-point for the parameters: $\Hessian_{xx}/B\OmgB=-\Hessian_{yy}/B\OmgB=1/10$,  $\OmgE^2=0$, and $\OmgD^4/\OmgB^4=-1/10^2$.  }
\label{fig:Orbits-Xpoint}
\end{figure}

Near a potential X-point, where $\Omega_D^4<0$, there is one stable and one unstable motion. 
Figure~\ref{fig:Orbits-Xpoint} illustrates the case of trajectories near a potential X-point.
If there are unstable motions, the theory of adiabatic invariants still applies in a qualitative manner. \cite{Best68physica}
However, as illustrated by  Fig.~\ref{fig:Orbits-Xpoint}, the real difficulty lies in the fact that the quadratic approximation will  break down because the orbit rapidly moves to  a region  that is far from the fixed point where the potentials may have a rather different dependance on space and time.

%%%%%%%%%%%%%%% 1D Surface %%%%%%%%%%%%%%%%

\section{Strong 1D Variation Across Magnetic Flux Surfaces \label{sec:1d_variation}}
\subsection{ Oscillation Frequency \label{sec:1d_GC_freq}}
  If the region far from a fixed point has strong variation along the magnetic flux coordinate,  $\xcoord:=\psi$, then there is single adiabatic invariant. 
  Expand the potentials via
 \begin{align}
 \Phi&= \Phi_0'\xcoord+ \Phi_0'' \xcoord^2/2 \\
  \VA&=\VA_0' \xcoord + \VA_0'' \xcoord^2/2=   B_0\xcoord \yhat   -  \mu_0\VJ_0  \xcoord^2/2.
 \end{align}  
   In this case, the effective potential is defined as
  \begin{align}
\DelPotential%&=     \nabla_\perp^2 \Phi  +\Vvdrift\cdot  \Vnabla\times \VB  =  \VB \cdot \Vnabla\times \Vvdrift   \\
  &:= \Phi_0'' -\VA''_0 \cdot  \Vvdrift_0=\Phi_0''+\mu_0 \VJ_0\cdot   \Vvdrift_0  
  \end{align}
 and determines the electric oscillation frequency via
 \begin{align}
 \OmgE^2  = \charge \DelPotential/\mass=\OmgB\partial_x V_{0,y}=  \OmgB\bhat \cdot\Vnabla\times\VV_0 .
 \end{align}

There is only a single pair of oscillation frequencies $\pm\OmgSlab$ where
  \begin{align}
\OmgSlab^2  &=\OmgB^2+ \OmgE^2= \OmgB^2+\OmgB\bhat \cdot\Vnabla\times\VV_0 .
  \end{align}
It is clear that near a potential minimum, $\charge \DelPotential>0$, the electric field gradient increases the oscillation frequency, whereas near a potential maximum, $\charge\DelPotential<0$, the electric field gradient decreases the oscillation frequency.   The orbits are stable for 
\begin{align}
\OmgE^2/\OmgB^2 = \DelPotential/\Bfield_0\OmgB > -1 .
\end{align}

The Taylor series for $\OmgSlab/\OmgB$, expressed as a function of $\OmgE^2/\OmgB^2$, only converges in the region $\norm{\OmgE/\OmgB}^2 <1$. Hence,  it is clear that standard GC and GK perturbation series do not converge outside of this region.

\subsection{Oscillatory Motion \label{sec:1d_GC_osc}}
The Hamiltonian in the guiding center reference frame can be written to second order as 
\begin{align}
\HamOsc=[\px^2+ (\py- \rx \charge \Bz)^2+\pz^2] /2\mass + \charge \DelPotential\xcoord^2/2 .
\end{align}
The lack of dependence of the Hamiltonian on $\ycoord$ and $\zcoord$ to this order implies that the momenta $\py$ and $\pz=\mass \vz$ 
are  conserved. 
 The equation  of motion for $\vx$
 \begin{align}
\mass \dot \vcoord_\rx &=\charge(\vcoord_\ry \Bz-\DelPotential\xcoord)
 \end{align}
can be solved by inserting $\mass\vcoord_\ry  =\py - \rx \charge \Bz$. This yields simple harmonic motion with frequency $\OmgSlab$ (Eq.~\ref{eq:OmgSlab}) and average position $\Xcenter$
  \begin{align}
 \dot v_\rx &= \OmgSlab^2(\Xcenter -\rx)
&
\Xcenter &=\py  \OmgB/\mass \OmgSlab^2.
\label{eq:Xcenter}
 \end{align}
The motion in $\ry$ is oscillatory, but with a different amplitude
  \begin{align}
     \vcoord_\ry &= (\Xcenter-\rx) \OmgB & \ry&= \vx \OmgB/\OmgSlab^2 .
 \end{align}
 
\begin{figure}[t]
\center
\includegraphics[width=3.5in]{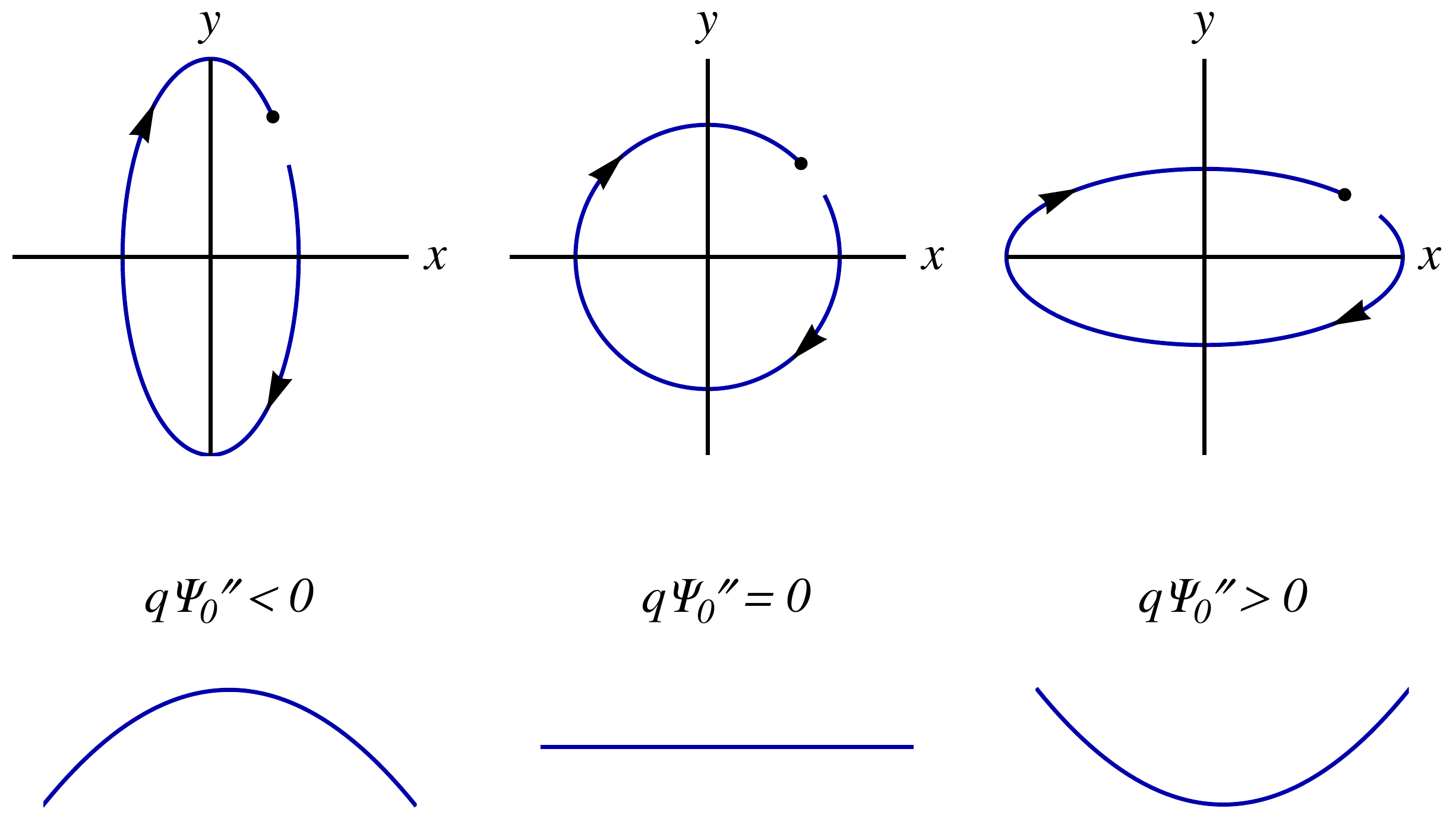}
\caption{Particle orbits become elliptical when the electric field is sheared: potential hill (left), flat potential (middle), potential well (right). Positively charged particles rotate clockwise for a magnetic field that points out of the page. }
\label{fig:Orbits-Slab}
\end{figure} 
 
%%%%%%%%%%%%%%%%%%%%%%%%%%%%%%
The gyro-orbit is given by
\begin{align} \label{eq:gyro_orbit_def}
\Vgyro&= \xhat \gyromag_\rx \sin{(\gyrophase)} -\yhat \gyromag_\ry \cos{(\gyrophase)} \\
\Vwcoord&= \xhat \wmag_\rx \cos{(\gyrophase)} + \yhat  \wmag_\ry \sin{(\gyrophase)} .
\end{align}
The angle $\theta$ is right-handed, but the rotation is diamagnetic; i.e. left-handed for a positively charged particle.
Thus, to lowest order,  $\Vw  =-\OmgSlab \partial_\gyrophase \Vgyro$ and $\Vwmag=\OmgSlab\Vgyromag$. 
However, due to the strong electric field, the oscillatory motion is elliptical rather than circular.  
The ratio of the principle axes of the ellipse  is determined via the relation
\begin{align}
\gyromag_\ry/\gyromag_\rx&=\wmag_\ry/\wmag_\rx =  \OmgB/\OmgSlab .
\end{align}
As shown in Fig. \ref{fig:Orbits-Slab}, the ellipse is longer in the $\rx$ direction for a potential well,  $\charge\DelPotential>0$, and is shorter in the $\rx$ direction for a potential hill,  $\charge \DelPotential<0$.  

The adiabatic invariant associated with the motion is  
\begin{align}
\Jmu= \mass (\vx^2+\OmgSlab^2 \rho_x^2)/2\OmgSlab= \mass \wmag_\rx^2/2 \OmgSlab =  \mass  \gyromag_\rx^2\OmgSlab /2 .
\end{align}
The second order Hamiltonian is simply $\HamOsc=\Jmu \OmgSlab$.

\subsection{Guiding Center Drifts \label{sec:1d_GC_drift}}
Now that the Hamiltonian has been computed, the guiding center drift equation can be refined to include the $\CO(\gyro^2)$ contribution 
\begin{align} \label{eq:guiding_center_forcebalance}
  \mass \dot  \VV +\mass \VV\cdot \Vnabla \VV +\Jmu\Vnabla\OmgSlab &=\charge  \VE + \charge\VV\times \VB  .
  \end{align}
  The zeroth order electric field $\VE_{0}$ drives the zeroth order drift $\Vvdrift_0=\VE_0\times\bhat_0/\Bfield_0$.
The first order drift is driven by the first order force
\begin{multline} \label{eq:first_order_force}
 \VF_{1} =\charge   \VE_1   +\VV_0\times\charge\VB_1 %\\
-    \partial_\tcoord \mass \VVcoord_0 -\mass V_0^i\VV_0\cdot \Vnabla \ehat_i -\Jmu  \Vnabla \OmgSlab .
\end{multline}
This expression is quite familiar from the standard case. 
Now, there are additional curvature forces $-V^i \VV\cdot \nabla \ehat_i$ because both $V^y$ and $V^z$ can be large.
In fact, the term $V^y V^y\partial_y \ehat_y\sim\CO(\epsilon)$ is dominant.

If one expands the velocity in terms of $\epsilon$ as $\Vvdrift=\sum_n \epsilon^n \Vvdrift_n$, then  $\VV_{n,\perp}$   is determined by an equation of the form
\begin{align}  
  \mass \Vvdrift_{n\perp}\cdot \Vnabla \Vvdrift_0  &= \VF_{n\perp}+  \Vvdrift_{n\perp} \times \charge\VB_0  
  \end{align}
 where $\VF_n$ denotes the sum of all additional ``force'' terms at order $n$. The solution for the drift
\begin{align} \label{eq:drift_x}
\vdrift_{n}^x&=F_{n,y}/\left(\charge \Bz + \mass \partial_x \vdrift_{0,\ycoord} \right) =F_{n,y}\OmgB/\mass \OmgSlab^2 \\
\vdrift_{n}^y&=- F_{n,x}/\charge\Bz =-  F_{n,x}/\mass \OmgB
\label{eq:drift_y}
\end{align}
is anisotropic in its response to the forces due to the additional inertia provided by the strongly sheared flow.
Thus the first order drifts are given by:
\begin{multline} \label{eq:first_order_drift_x}
\vdrift_{1}^x =\left( \charge \Efield_{y,1}   +  V^z\charge\Bfield_{x,1}    -\Jmu  \partial_y \OmgSlab   \right.\\
 \left.- \yhat \cdot  \partial_\tcoord  \mass\VVcoord_0 -\mass V_0^i \yhat \cdot ( \VV_0\cdot \Vnabla ) \ehat_i \right)\OmgB/\OmgSlab^2 
\end{multline}
and
\begin{multline}
\vdrift_{1}^y =\left( -\charge\Efield_{x,1}   -  V^y\charge\Bfield_{z,1}+   \charge V^z\Bfield_{y,1}  +\Jmu  \partial_x \OmgSlab   \right.\\
 \left.+ \xhat \cdot  \partial_\tcoord  \mass\VVcoord_0 +\mass V_0^i \xhat \cdot ( \VV_0\cdot \Vnabla )\ehat_i \right)/\OmgB.
\end{multline}

\def\DensityGC{\Density_\mathrm{gc}}
\def\Pressure{P}
\def\PressureGC{\Pressure_\mathrm{gc}}
The time derivative of the drift flow $d\VV_{n,\perp}/d\tcoord$ generates a polarization drift and polarization charge density
at one higher order in $\epsilon$:
\begin{align} \label{eq:drift_pol}
\Vvdrift_{n+1,\mathrm{pol}}&= \frac{d}{d\tcoord}\left(  \frac{\VF_{n,\perp}}{\mass \OmgSlab^2}\right)  
\\
\Density_{n+1,\mathrm{pol}}&= -\Vnabla \cdot  \frac{\DensityGC \VF_{n,\perp}}{\mass \OmgSlab^2}.
\end{align}
Interestingly enough, the polarization  is gyrotropic, i.e. isotropic in the directions perpendicular to the field.

\subsection{Polarization and Magnetization \label{sec:1d_GC_pol} }
The  polarization and magnetization can be computed in a straightforward manner following the
microscopic ``bottom-up'' approach.  \cite{Brizard07rmp, Brizard13pop}
The intrinsic microscopic polarization vector $\Vpi$  and  macroscopic polarization  density $\VPolarization$ in guiding center space are defined by 
\begin{align}     \label{eq:pol_def}
\Vpolarization  &=\charge   \int \left[\Vgyro -\tfrac{1}{2} \Vnabla \cdot \Vgyro\Vgyro+\dots\right]d\gyrophase/2\pi
\\
\VPolarization   &= \int \Vpolarization \Pdf d\Jmu \OmgSlab d\vpar 
   \label{eq:Pol_def}
\end{align}
where $\Pdf$ is the guiding center distribution function.

For any force $\VF$, the time derivative of the flow introduces a polarization drift. The finite displacement for $\avg{\Vrho}=\oint dt \Vcoord_{2,\mathrm{pol}}$  is derived from the  polarization drift, Eq.~\ref{eq:drift_pol}, which leads to
\begin{align}
\avg{\Vrho} =  \VF_\perp/\mass\OmgSlab^2.
\end{align}
 Thus, the result for the guiding center polarization is
 \begin{align}    
 \label{eq:pol_result}
\Vpolarization  &=\charge     (\charge\VE_\perp-\Jmu\Vnabla_\perp \OmgSlab)/\mass\OmgSlab^2 
-  \tfrac{1}{2}\Vnabla  \cdot   \charge     \left[\gyromag_\rx^2\xhat \xhat+ \gyromag_\ry^2 \yhat\yhat  \right] 
\\
\VPolarization  &=  \charge(\charge  \DensityGC \VE_\perp-\Pressure_{\mathrm{gc},xx} \Vnabla_\perp \ln{(\OmgSlab}))/\mass\OmgSlab^2  
  -  \Vnabla  \cdot  \charge \pressgctensor_{\mathrm{gc},\perp} /2\mass\OmgSlab^2  
 \label{eq:Pol_result}
\end{align}
where $\pressgc_{\mathrm{gc},ij}$ is the zeroth order guiding center pressure tensor.

Note that when comparing to magnetized fluid theory, it is important to recognize that the guiding center density, $\DensityGC$, itself differs from the particle density, $\density_0$, in the limit of vanishing gyroradius. As explained in Ref.~\onlinecite{Joseph19gkpol}, the difference $\DensityGC-\density_0$, contributes an amount that is equal to the pressure term in Eq.~\ref{eq:Pol_result}. Accounting for the contribution of both terms doubles the magnitude of the diamagnetic polarization in the adiabatic drift-reduced fluid theory. 

The intrinsic microscopic magnetization vector $\Vmagnetization$ and macroscopic magnetization density $\VMagnetization$ are defined by
\begin{align}\label{eq:mag}
\Vmagnetization  &= 
 \charge \int\left[\Vgyro\times \VV_\perp + \tfrac{1}{2}\Vgyro \times \tfrac{d}{d\tcoord} \Vgyro+\dots\right]d\gyrophase/2\pi
\\%&
\VMagnetization &=  
\int \Vmagnetization  \Pdf d\Jmu \OmgSlab d\vpar
\label{eq:Mag}
\end{align}
where the first term in Eq.~\ref{eq:mag} is the moving dipole term due to the electric polarization. The results for the magnetization
\begin{align}
\Vmagnetization  &= 
-\charge (\Jmu\OmgSlab+\mass  \Vcoord_y^2)  \OmgB \bhat/\mass\OmgSlab^2
\\%&
\VMagnetization &=  
-\charge (\Pressure_{\mathrm{gc},xx}+\DensityGC \mass  \Vcoord_y^2)    \OmgB \bhat/\mass\OmgSlab^2
\end{align}
are modified by the change in frequency and by the addition of the zeroth order ram pressure.

\section{Extension to Gyrokinetic Orbit Theory \label{sec:GK}} 
GK theory extends the limits of GC theory to arbitrary $\kperp\gyro$, where $\kperp$ is the perpendicular wavenumber. However, this assumes that the amplitude of the variation is sufficiently small.  Consideration of the particle motion near the potential minimum of a sinusoidal wave implies   that the restriction requires
\begin{align}
\delta_E=\kperp |E_{\perp,\kcoord}|/\OmgB\Bfield \lesssim 1
. 
\end{align}
This restriction is satisfied by both the large flow $\delta_E\sim\CO(\epsilon)$  and  standard  $\delta_E \sim \CO(\epsilon^2)$ orderings. \cite{Dubin83pof,Hahm96pop,Dimits10pop}

The GC theory for strong variation across magnetic flux surfaces, presented in Sec. \ref{sec:1d_variation}, can readily be used to develop a more accurate GK theory
 for the motion of gyrocenters.
The main differences are due to the fact that the drift motion is anisotropic and that the orbits are elliptical.  Given the convention adopted in Eq.~\ref{eq:gyro_orbit_def}, the quantity $\Vkcoord\cdot \Vgyro=\kdotrho  \sin{(\gyrophase-\gyrophase_\kcoord)}$ has the magnitude and reference angle:
\begin{align}
\kdotrho^2 &=(\Vkcoord \cdot \Vgyromag)^2=(\kcoord_\rx \gyromag_\rx)^2+(\kcoord_\ry \gyromag_\ry)^2\\
\tan{(\gyrophase_\kcoord)} &=  \kcoord_\ry \gyromag_\ry/\kcoord_\rx \gyromag_\rx= \kcoord_\ry \OmgB/\kcoord_\rx \OmgSlab
.
\end{align}
Many of the formal expressions that appear in the GK theory are identical with the replacement of the  frequency $\OmgB\rightarrow \OmgSlab$ and the expressions above for $\kdotrho$ and $\gyrophase_\kcoord$. 
However, the drifts must be modified to the anisotropic  form given in Eqs.~\ref{eq:drift_x}-\ref{eq:drift_y}.

\def\BesselG{G}
The perturbed first-order electric potential $\Phipert(r)$ in guiding center coordinates is defined via
\begin{align}
\Phipert(\Rcoord,\mu,\gyrophase)&=\sum_\kcoord \int dr e^{ i\Vkcoord\cdot (\VR-\Vr)} \Phi_1(r)/(2\pi)^3 \\
&=\sum_\kcoord \int dr e^{-i\Vkcoord\cdot \Vgyro} \Phi_{1k} /(2\pi)^3.
\end{align}
Hence, the potential in real space can be expanded as 
\begin{align}
\Phipert(r)&=\sum_{ \ncoord } \sum_\kcoord e^{i\ncoord(\gyrophase-\gyrophase_\kcoord)+i\Vkcoord \cdot \VRcoord}\BesselJ_\ncoord(\kdotrho)\Phi_{1\kcoord}
\end{align}
where $\BesselJ_n(x)$ is the Bessel function of order $n$.
The contributions to the Hamiltonian from the  vector potential that are of the form  $\VV\cdot \VA_1  $ and $\vpar A_{1\|} $ can be expressed in a similar fashion.
The contribution of the perpendicular component $(\Vv-\VV)\cdot\VA_{1\perp}$ can be written in terms of $\Bfield_{1\|}:=\bhat\cdot\Vnabla \times \VA_1$ as
\begin{multline}
(\Vv-\VV)\cdot\VA_{1\perp}(\Vr) =\sum_{ \ncoord } \sum_\kcoord e^{i\ncoord(\gyrophase-\gyrophase_\kcoord)+i\Vkcoord \cdot \VRcoord} \times
\\
	\left[\BesselJ_{\ncoord+1}(\kdotrho)-\BesselJ_{\ncoord-1}(\kdotrho)\right] \mu \Bfield_{1\|\kcoord} \OmgB/2\kdotrho\OmgSlab.
\end{multline}
Here, the magnetic moment, $\mu = J\charge/\mass$, is used to represent the adiabatic invariant in order to eliminate any possible confusion with the Bessel functions.

\def\opAccent{\hat}
\def\kdotrhohat{\opAccent\krho}
\def\Operator{O}
The first order GK gyrocenter Hamiltonian is simply the $n=0$ orbit average of these quantities. 
The  two required gyroaveraging operators are defined by $\BesselJ_0(\kdotrho)$ and 
\begin{align}
\BesselG_1(\kdotrho)=\BesselJ_1(\kdotrho)/\kdotrho=\left[\BesselJ_0(\kdotrho)+\BesselJ_2(\kdotrho)\right]/2.
\end{align}
Each averaging operator, defined by its Fourier expansion, $\Operator(k)$, acts on a function, $\pdf(R)$, as an integro-differential convolution operator, $\opAccent\Operator$, via
\begin{align}
\opAccent \Operator f &:=\int d\Rcoord' \sum_\kcoord e^{i\Vkcoord \cdot (\VRcoord-\VRcoord')}\Operator(k) f(\Rcoord')/(2\pi)^3 .
\end{align} 
The first order GK gyrocenter Hamiltonian is
\begin{align}
\HamGK &=  \charge\left( \opAccent \BesselJ_0 \Phi_1 -   \VV\cdot \opAccent  \BesselJ_0 \VA_{1 \perp } - \vpar  \opAccent  \BesselJ_0 A_{1\|} \right) +\mu \opAccent \BesselG_1    \Bfield_{1\|} \OmgB/\OmgSlab .
\end{align}
If one assumes that the perturbation amplitudes are of order $\delta$, where $\epsilon \ll\delta\ll 1$, then the total Hamiltonian is 
\begin{align}
\Hamiltonian=\HamDrift+\HamOsc + \HamGK.
\end{align}
The gyrokinetic drifts are anisotropic and result from applying Eqs.~\ref{eq:drift_x}-\ref{eq:drift_y} to the first order Hamiltonian and momentum.

%%%%%%%%%%%%%%% Conclusion %%%%%%%%%%%%%%%%
\section{Conclusion \label{sec:conclusion}}
In conclusion,    guiding center (GC) and gyrokinetic (GK) theory can be extended to the regime of large  electric field gradient, which directly modifies the oscillation frequency and causes the orbits to become elliptical. 
In order for the trajectories to retain  a robust adiabatic invariant and a slower drift motion, the  spatial variations can only depend strongly on a single coordinate at lowest order. 
For example, the present theory would  apply to  shear flows that are even $10\times$ larger than those
observed in the H-mode  pedestal even if the toroidal/guide magnetic field is as small or even smaller than the poloidal field. 
The resulting GC/GK theory displays anisotropy in the drift motion and modifies the polarization and magnetization, but otherwise retains a similar mathematical form to the standard case. 
Thus, the changes needed to improve the accuracy can readily be implemented in existing GC/GK simulation tools.  
It is of great interest to continue  exploring the  physical implications of the extended ordering, % and 
its convergence for strong field  variations, and to develop the fully self-consistent set of GC/GK Maxwell's equations.  
It would also be valuable to further develop the connection to oscillation center theory and the theory of waves in magnetized plasmas.

\appendix

\begin{acknowledgements}
The author  would like to thank D.~D.~Ryutov and A.~M.~Dimits for inspiring research on this topic and 
M.~A.~Dorf for insightful discussions that led to an improved understanding of the anisotropic character of the drift motion. 
The author would also like to thank B.~I.~Cohen and M.~A.~Dorf for carefully reading an early version of the manuscript.
This work, LLNL-JRNL-813195, was performed under the auspices of the US DOE by LLNL under Contract DE-AC52-07NA27344.
\end{acknowledgements}

\section*{Data Availability}
The data that support the findings of this study are available from the corresponding author
upon reasonable request.

\bibliography{GC_Efield_POP_v5}% Produces the bibliography via BibTeX.

\end{document}